# Inline Quantum Measurements with SNSPDs Coupled to Photonic Bound States

Filippo Martinelli[1,2], Anton N. Vetlugin[1,2], Shuyu Dong[2,3], Darren M. Z. Koh[4], Mariia Sidorova[2,†], Christian Kurtsiefer[4,5], Cesare Soci[1,2,3,*]

*[1]Division of Physics and Applied Physics, School of Physical and Mathematical Sciences, Nanyang Technological University, Singapore 637371*

*[2]Centre for Disruptive Photonic Technologies, TPI, Nanyang Technological University, Singapore 637371*

*[3]School of Electrical and Electronic Engineering, Nanyang Technological University, Singapore 639798*

*[4]Centre for Quantum Technologies, National University of Singapore, Singapore 117543*

*[5]Department of Physics, National University of Singapore, Singapore 119077*

*[†]Current affiliations: Humboldt-Universität zu Berlin, Department of Physics, Newtonstr. 15, 12489 Berlin, Germany; German Aerospace Center (DLR), Institute of Optical Sensor Systems, Rutherfordstr. 2, 12489, Berlin, Germany*

* csoci@ntu.edu.sg

**Abstract:** We report the realization of inline quantum measurements with waveguide-integrated superconducting nanowire single-photon detectors (SNSPDs). To suppress parasitic scattering at detector terminations, while ensuring compatibility with standard photonic substrates and cryogenic operation, we developed a photonic bound states in the continuum (BIC) platform based on etchless polymer waveguides. We show BIC-coupled inline detectors with on-chip efficiency exceeding 80%, recovery time of less than 2 ns, and intrinsic jitter of less than 70 ps. As a proof of principle, we implement Hanbury Brown and Twiss interferometry and photon number resolution with two collinear detectors within a footprint of 60×6 μm$^2$. The concept of inline quantum measurements could be further developed to support more complex circuit functionalities, such as higher-order correlation measurements, quantum state tomography, and multi-photon subtraction, within a compact architecture.

## 1. Introduction

Rapid advancements in integrated photonics have enabled the scaling and miniaturization of quantum circuits, paving the way for complex experiments that would be impractical with conventional bulk optical setups [1-4]. A central component in these circuits is the single-photon detector [5, 6], which performs the critical task of measuring the quantum state. Among available technologies, superconducting nanowire single-photon detectors (SNSPDs) stand out for their high detection efficiency [7], excellent timing performance [8], minimal noise [9], and relatively straightforward fabrication [5]. This combination of performance and versatility has enabled their successful integration into a wide range of photonic platforms, including silicon-on-insulator [7], silicon nitride [10-12], and lithium niobate [13-15] waveguides and circuits.

A distinctive feature of integrated SNSPDs is the ability to tailor the absorption efficiency by controlling the length of nanowire coupled to the waveguide [16]. Although this has typically been exploited to maximize absorption at the termination of photonic circuits, this tuneable coupling can serve a broader purpose [4, 17, 18]. In this work, we introduce and demonstrate the concept of inline detection using integrated SNSPDs. By cascading multiple nanowires with controlled, partial absorption along a single waveguide, we show that complex detection functionalities can be achieved within a compact footprint. To illustrate this approach, we implement an integrated Hanbury-Brown and Twiss [19] (HBT) interferometer and a photon number resolving (PNR) detector, two essential tools for quantum light detection and characterization [20, 21]. Specifically, we design and fabricate an inline HBT interferometer consisting of two SNSPDs integrated along a single waveguide and validate its ability to perform photon correlation measurements. We also demonstrate that the same approach can be used to reconstruct the photon statistics by distributing photon absorption probability across different wires. We conclude by discussing the broader implications of inline detection with SNSPDs for scalable quantum photonic circuits.

## 2. Results

To enable the integration of multiple inline detectors, we developed an unconventional photonic circuit architecture based on photonic bound states in the continuum (BIC) [22]. In this design, superconducting nanowires are embedded beneath polymer ridges that confine light within the silicon substrate – departing from conventional silicon photonic chips where light is guided in high-index waveguide core elevated from the substrate.

This architecture decouples the design of the SNSPD electronic readout circuitry in the bottom layer from the optical waveguides defined in the top layer, mitigating existing limitations of conventional integrated SNSPDs, most notably the scattering losses at the crossing structures at the detector terminations. These structures are known sources of scattering losses [23], which would be particularly detrimental in inline configurations, where multiple detectors are cascaded along the same waveguide. Unlike previous demonstrations [18], our method does not compromise single mode operation allowing inline measurements in compact photonic circuits.

Beyond reducing parasitic losses, the use of photonic bound states eliminates the need for etching steps, thereby simplifying fabrication. We demonstrated detectors integrated with BIC waveguides and benchmarked their performance against standard approaches, showing field confinement and propagation losses comparable to conventional silicon and silicon nitride waveguides, even at cryogenic temperatures.

Using this platform, we implemented an inline Hanbury Brown and Twiss (HBT) interferometer to perform autocorrelation measurements of a heralded single-photon source. Furthermore, we demonstrated the PNR capabilities of two inline detectors by measuring the photon statistics of coherent states and comparing the results with theoretical predictions.

### *2.1 Photonic bound state waveguides*

A thin dielectric layer cannot confine and route light, as it supports a continuum of unbound optical waves, rather than a discrete set of well-confined modes [24]. However, a strip of a different material, placed on top of the dielectric layer, creates an effective potential well that can confine discrete transverse magnetic (TM or quasi-TM) or transverse electric (TE or quasi-TE) modes, **Fig. 1**. Among these, TM modes exhibit a stronger interaction with the nanowire, making them a preferrable choice for the design of compact and high-speed integrated detectors (**Fig. 1a**).[25]

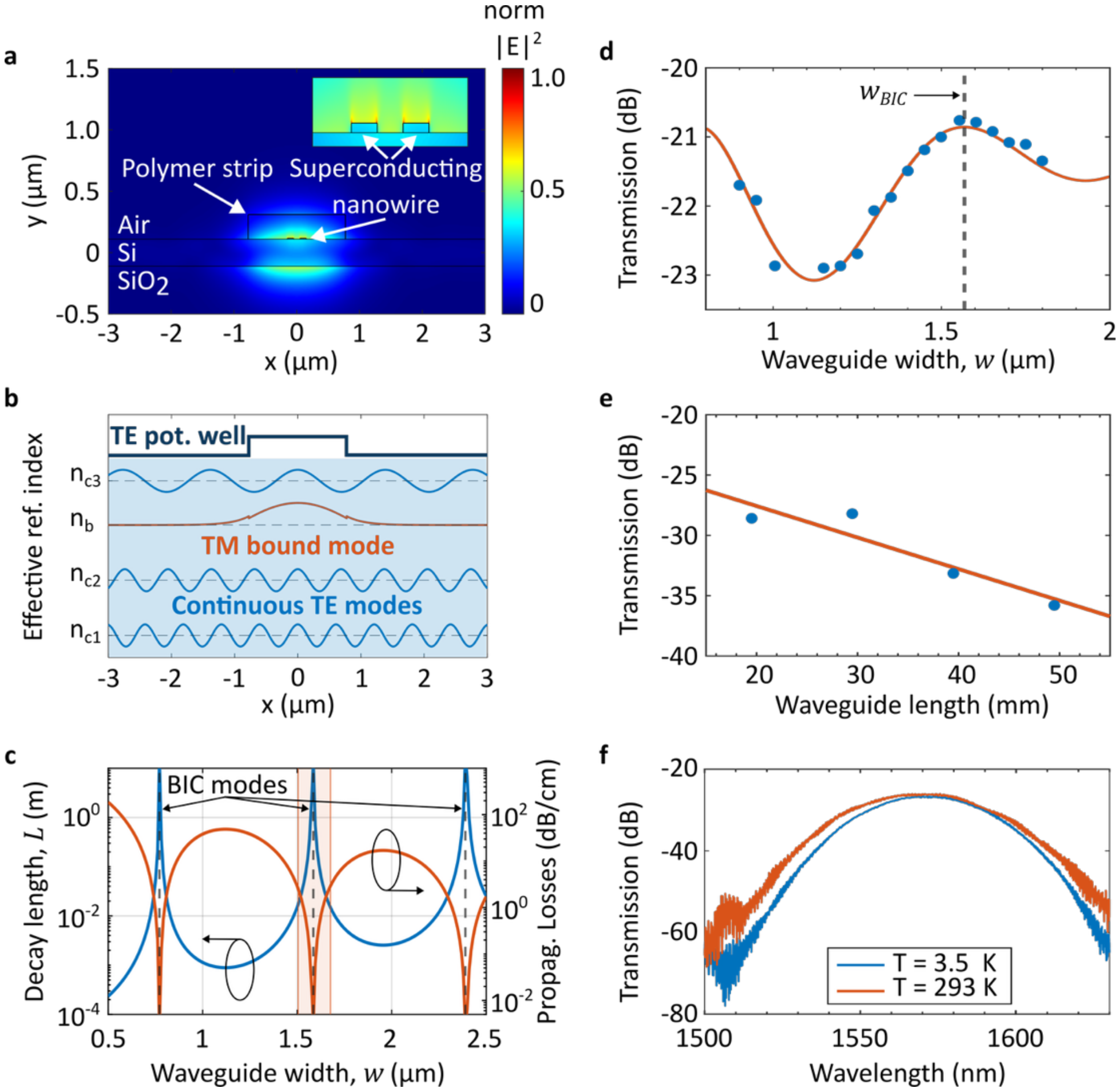

**Figure 1. Photonic bound state waveguides for inline quantum light detection**. (a) Cross sectional view of the electric field intensity of the confined TM mode around the integrated nanowire detector[26]. The inset shows a zoom-in of the nanowire region. (b) Schematic representation of the effective refractive index and spatial distribution of a few continuous TE modes (effective refractive indices $n_{c1}$, $n_{c2}$, and $n_{c3}$) and bounded TM mode ($n_b$). (c) Calculated decay length (blue curve) and propagation losses (orange curve) of the TM mode as a function of the waveguide width. The orange-shaded interval marks the region with propagation loss below 3 dB/cm. (d) Transmission as a function of the width for 1 mm long waveguides. $w_{BIC}$ shows the peak of the transmission as calculated from the fitting. (e) Transmitted power as a function of the BIC waveguide length. (f) Waveguide transmission spectra at T = 3.5 K (blue) and room temperature (orange).

As illustrated in **Fig. 1b**, the effective index of the TM mode ($n_b$) lies within the continuum of TE slab modes. Although one might expect absence of coupling between modes of opposite polarization, it has been shown that significant interaction can occur between the TM waveguide mode and the TE slab modes at the core-cladding interface [27]. Under total internal reflection, the incident TM mode is partially coupled into two TE slab modes, one reflected in the core region and the other transmitted in the cladding region [28]. Because the slab modes can propagate in arbitrary directions, there exists a specific propagation angle at which they are phase-matched to the TM waveguide mode. As a result of this coupling, the bound TM mode gradually loses power to the phase-matched slab modes.

The characteristic decay length of the TM mode, $L(w)$, depends on the waveguide width $w$ and obeys a periodic behaviour [29] (**Fig. 1c**):

$$L(w) = \frac{4L_0}{k_x^2}\left(\mathrm{sinc}\frac{k_x w}{2}\right)^{-2}. \tag{1}$$

Here, $k_x$ is the transverse component of the wavevector of the phase-matched TM (discrete) and TE (from the continuum) modes and $L_0$ is the coupling strength between them. For certain waveguide widths, integer multiples of $\frac{2\pi}{k_x}$, the decay length diverges, indicating a suppression of the TM dissipation. This corresponds to a condition of destructive interference between the two slab modes generated at the waveguide interface. In fact, the radiation leaking from the waveguide to the substrate is the coherent sum of the reflected and transmitted TE modes at the two waveguide interfaces and depends strongly on their relative phase [28]. With a proper choice of the waveguide width, this interference can become fully destructive, suppressing the coupling to the continuum and forming a bound state in the continuum [29-31]. These states are marked by vertical dashed lines in Fig. 1c.

We implemented the BIC waveguide on a silicon-on-insulator (SOI) platform with 220 nm silicon layer on top of a silicon dioxide substrate (fabrication details are available in Methods). As a dielectric strip defining the guided TM mode, we used polymer CSAR 62, which can operate at cryogenic temperatures [32] and can be processed with a single lithography step. The height of the polymer has negligible influence on the BIC position and was set at 200 nm. The photonic circuit was optimized for operation at 1550 nm, as C-band photonic qubits are naturally suited for photonic quantum technologies due to the maturity of telecommunication components [4]. We chose to operate at the BIC mode occurring at the waveguide width of 1.58 μm, which allows a small footprint (compared to higher-order modes) and, at the same time, is tolerant to fabrication imperfections (compared to the lower-order modes). In support of the latter, we estimated that the propagation loss is kept beyond -3 dB/cm for a $\pm$5% uncertainty of the waveguide width (the orange shaded area in Fig. 1c), which is within the tolerance of our current waveguide fabrication process.

To identify the BIC condition experimentally, we fabricated waveguides with a fixed length of 1 mm and varying widths ranging from 0.8 to 1.8 μm. Light transmission $T_{\mathrm{dB}}(w)$ through such waveguides, according to Eq. (1), should exhibit a sinc-squared behaviour:

$$T_{\mathrm{dB}}(w) = 2C_{\mathrm{gc}} - \frac{5lk_x^2 \log_{10} e}{2L_0}\left(\mathrm{sinc}\frac{k_x w}{2}\right)^2, \tag{2}$$

where $l$ is the length of the waveguide. We accounted for coupling in and out of the waveguide with the coupling efficiency of $C_{\mathrm{gc}}$ expressed in dB (see below), under the assumption of symmetric in-coupling and out-coupling. Considering that in the range of waveguide widths of interests the effective refractive index variation is less than 2%, $k_x$ is taken as constant and transmission in Eq. (2) depends only on the waveguide width $w$. Measurement results (blue dots) presented in **Fig. 1d** agree with the predicted behaviour and are well fitted by Eq. (2). From the fitting, we located the BIC regime at the waveguide width of approximately 1.57 μm, which agrees with the numerical simulation results within a 1% error.

To verify that the chosen waveguide width leads to suppression of the guided mode dissipation, we characterized the propagation losses in the BIC waveguide. **Figure 1e** shows the measured transmission through waveguides of different length and optimal width of 1.57 m. The slope of the transmission curve corresponds to a -2.6 dB/cm loss rate in the waveguide, assuming identical coupling efficiency among different samples. We attribute this propagation loss rate to the 5% fabrication uncertainty discussed above. Further reduction of propagation losses may be obtained by optimizing the fabrication process, specifically by achieving higher control on the waveguide width and reducing the polymer sidewall roughness.

In view of the following integration of superconducting detectors, we also verified the robustness of polymer-based waveguides at cryogenic temperatures. For this purpose, we compared the transmission spectrum through the waveguide at room temperature and 3.5 K (**Fig. 1f**) (details on the cryogenic setup are available in Methods). While, generally, the spectral response of the waveguides varied significantly at low temperature, likely due to a combination of the change of the polymer refractive index and variations in fibre-to-chip alignment in the cryogenic setup [33], the overall circuit performance was comparable in the target region of 1550 nm. We also observed that the performance of the waveguides did not degrade after at least five cooldown and warmup cycles.

These results confirm that the BIC waveguides are a reliable and robust platform for guiding light at cryogenic temperatures, enabling a new approach for superconducting detector integration.

### *2.2 Integration of superconducting detectors with photonic bound state waveguides*

The simple geometry and fabrication process of SNSPDs have enabled their successful integration with various photonic platforms, including silicon-on-insulator [7], silicon nitride [10-12], and lithium niobate [13-15]. SNSPDs are most commonly fabricated on top of photonic circuits [7, 10–15]. While there has been proposals to position the detector beneath the waveguide [34], practical implementations have proven challenging. In our platform, the use of polymer-defined BIC waveguides allows to reverse the integration sequence in a simple and practical manner: the SNSPDs are first fabricated on a flat silicon substrate, after which the polymer waveguide is patterned directly above them (see Methods for fabrication details).

This inverted architecture relaxes the requirement for the nanowire to strictly follow the waveguide core, as in previous demonstrations, and provides the flexibility to route it into and out of the core without additional supporting structures with scattering losses as low as 1.3 % (**Fig. 2,** and **Supplementary Information S2**). Since no bridge or support is required at detector crossings or terminations to electrically connect the nanowire to the readout circuit, scattering losses are reduced compared to conventional ridge waveguide designs without sacrificing footprint (see **Supplementary Information S3**). Thus, this feature makes the proposed BIC architecture well-suited for inline detection schemes.

We used niobium titanium nitride (NbTiN) SNSPDs for integration, selected for their combination of high detection efficiency, excellent timing resolution, and compatibility with operation at relatively high cryogenic temperatures (2.5-4.0 K) [35, 36]. The detectors employed a U-shaped geometry (**Fig. 2a**) with a nanowire width of 60 nm and a thickness of 7 nm. To characterize individual integrated detectors, we used a four-port waveguide circuit (**Fig. 2b**), where the outer loop between Ports 1 and 4 provided a reference to align a fibre array mounted on a four-axis nanopositioning stack, and the detectors were placed in the inner loop between Ports 2 and 3.

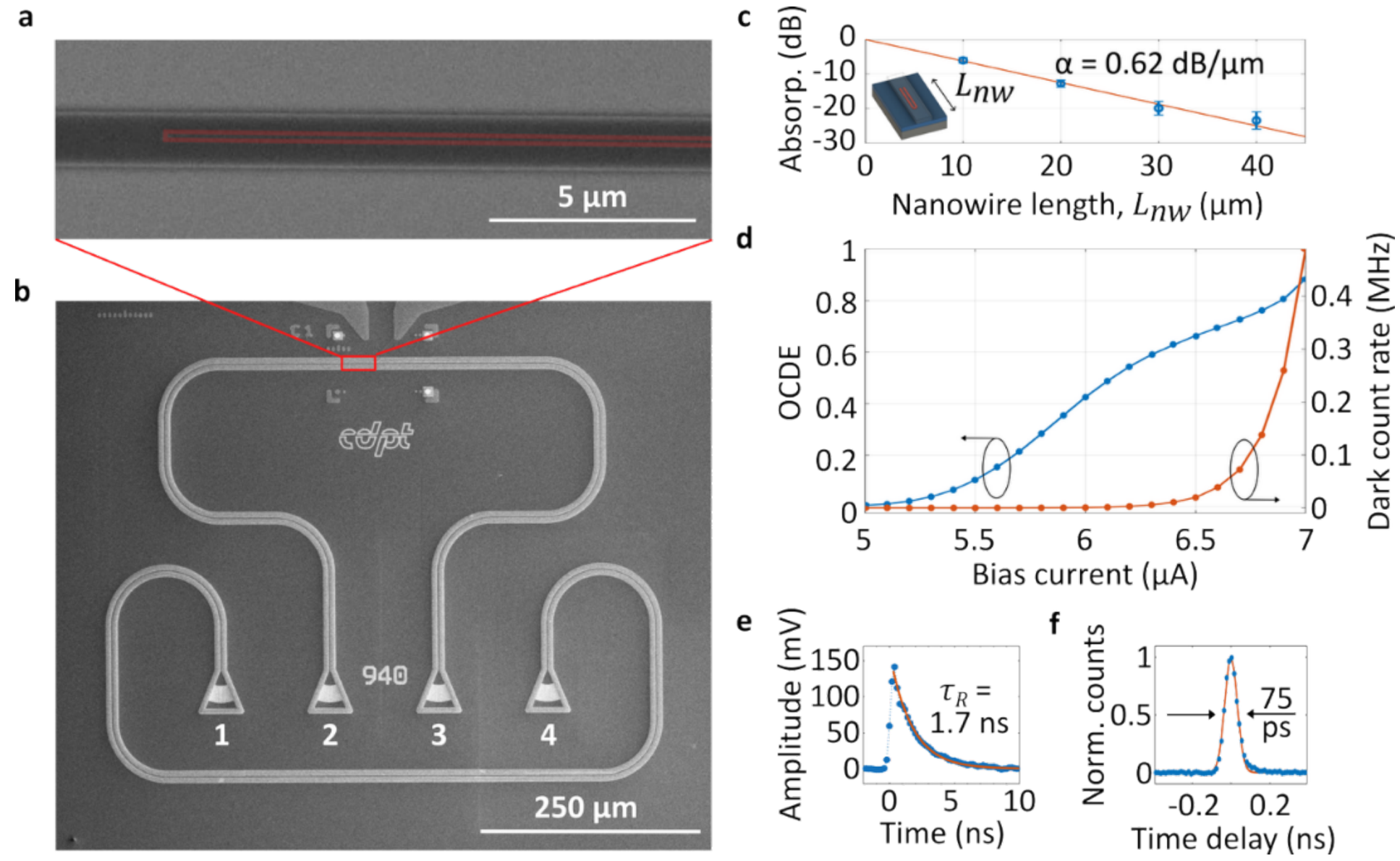

**Figure 2. Characterization of superconducting detectors integrated with photonic bound state waveguides.** (a) SEM image of the nanowire (red-coloured) embedded under the waveguide. (b) Microscope image of the 4-port BIC waveguide for integrated detector characterization. (c) Light absorption for different nanowire lengths. Linear fit of the data yields an absorption coefficient $\alpha = 0.62\ dB/\mu m$. (d) On-chip detection efficiency (blue) and dark count rate (orange) for a 45 µm long nanowire measured at 3.5 K. (e) Averaged output pulse waveform (dot) and exponential fitting (solid line). The recovery time calculated from the fitting is $\tau_R = 1.7$ ns. (f) System jitter histogram (blue dots) and its Gaussian fit (orange line). Arrows indicate the FWHM of the Gaussian fitting.

A critical parameter for integrated SNSPDs, particularly for the design of inline detectors, is the optical absorption rate (absorption per unit nanowire length). A higher absorption rate allows photons to be absorbed over shorter nanowire lengths, which benefits the detector's count rate, timing jitter, and fabrication yield. To characterize the absorption rate of BIC-integrated SNSPDs, we fabricated devices with varying nanowire lengths and measured their absorption at room temperature. In these measurements, a continuous-wave (CW) laser at 1550 nm wavelength was coupled into one of the inner waveguide ports and collected at the opposite port. **Figure 2c** shows the measured absorption as a function of nanowire length, from which we extracted the absorption rate of –0.62 dB/µm. Therefore, near-unity absorption efficiency (>99%) can be achieved with nanowires as short as 32 µm, comparable to that of SNSPDs integrated in conventional photonic platforms. Based on these results, we fixed the nanowire length to 45 μm for subsequent measurements with single SNSPDs, which is sufficient to ensure full (~99.9%) light absorption.

To estimate the on-chip detection efficiency (OCDE), we illuminated the detectors with continuous-wave (CW) laser attenuated to the single-photon level, while recording the detector count rate. Details on the estimation of photon flux and the coupling efficiency used in the OCDE analysis are provided in Methods. All the detector characterization was performed at 3.5 K. As shown in **Fig. 2d**, the OCDE exceeded 80% when the detectors were biased near the critical current. The absence of a plateau in the efficiency curve indicates that the internal quantum efficiency did not reach full saturation, suggesting potential to further increase the OCDE, for instance, by lowering the temperature to 2.5 K [37]. The orange curve in Fig. 2d shows the dark count rate (DCR) measured with the laser switched off. As expected, the DCR increases exponentially as the bias current approaches the critical value, consistent with prior observations in standard SNSPDs and attributed to the probabilistic hopping of magnetic vortices across the nanowire. [38]

Next, the detector recovery dynamics were investigated by analysing the output pulse, acquired with a 1 GHz oscilloscope after the amplification stage. **Figure 2e** shows the averaged output waveform, obtained from 16 consecutive acquisitions. The recovery time, defined as the decay constant of the exponential fit to the falling edge, was measured to be 1.7 ns, indicating the potential for count rates exceeding 100 MHz.

Finally, we characterized the timing jitter using start-stop histogram measurements with a reference femtosecond pulsed laser (see Methods). As shown in **Fig. 2f**, the full width at half maximum of the histogram was measured to be 75 ps. Accounting for the contribution from electronic and setup noise, the intrinsic jitter was estimated to be 69 ps, well within the expected range for this type of detectors and confirming that the new integration platform does not compromise timing performance.

Taken together, the results reported in this section demonstrate the viability of integrating NbTiN-based SNSPDs with BIC polymer waveguides. On-chip detection efficiency and recovery time of our integrated SNSPDs match the best parameters reported for the standard platforms [7, 39]. Importantly, this level of performance was achieved without compromising functionality, enabling the BIC platform to extend the capabilities of SNSPDs through the implementation of inline detection schemes, as presented in the following section.

*2.3 Inline quantum measurements*

To demonstrate the concept of inline detection, we designed and realized an integrated Hanbury Brown and Twiss interferometer [19] for on-chip autocorrelation measurements, and a photon number resolving detector for validating photon statistics.

Our inline implementation for autocorrelation measurements is presented in **Fig. 3**. Such measurements are commonly used in the characterization of quantum emitters, as they provide information about photon correlations [20] and indistinguishability [40]. **Figure 3a** illustrates the inline HBT setup, where two SNSPDs were placed sequentially along the waveguide.

The inset graph summarizes the design principle of the inline setup: the first detector, designed with a shorter nanowire, absorbs 50% of the incoming light, while the remainder is transmitted to the second detector. The second, longer detector achieves unitary absorption efficiency. This configuration effectively functions as a 50:50 beamsplitter with integrated detectors, replicating the behaviour of a conventional HBT interferometer based on a beamsplitter. The optical interferometer footprint occupied an area of 60×6 $\mu m^2$ (dashed box in Fig. 3a), defined by the total length of the nanowires (~60 µm in this realization) and the distance within which 99.9% of the optical mode is confined (~6 µm from the simulation in Fig. 1a) [41, 42].

As a proof of principle, we used this configuration to characterize the correlation properties of a heralded single-photon source based on spontaneous parametric down-conversion (SPDC) [20]. Unlike standard quantum emitters, heralded sources require the use of a conditional version of the second-order autocorrelation function, $g_c^{(2)}(\tau)$, which is evaluated with respect to the detection of a heralding photon [20, 42, 43]. The formal definition and details of this function are provided in Methods.

Photon pairs were generated in SPDC by pumping a periodically poled potassium titanyl phosphate (PPKTP) crystal with a CW laser at 532 nm. The idler photon, centred at 810 nm, was separated from the signal with a dichroic mirror and detected with a Si SPAD heralding detector (**Fig. 3b**). The signal photon, at 1550 nm, was coupled to the photonic chip and directed to the inline HBT setup for autocorrelation measurements. The conditional autocorrelation analysis requires monitoring triple coincidences among the heralding detector and the two inline SNSPDs, as well as the double coincidences (cross-correlation) between the heralding detector and each of the two inline SNSPDs.

The normalized cross-correlation between the heralding detector and the second inline SNSPD (D2), is shown in **Fig. 3c.** The sharp peak at zero delay, characteristic of the SPDC source, confirms the detection of correlated photon pairs. The curve shows a slightly asymmetric shape inherited from the response function of the heralding Si SPAD. As the jitter of the integrated SNSPD is 75 ps and the typical coherence time of the source is of the order of a few ps [44], the correlation function is dominated by the response function of the silicon heralding detector [45]. Fitting with an exponentially modified gaussian function yields a FWHM of ~460 ps, corresponding to a Si SPAD response function of ~450 ps. The cross-correlation between the heralding detector and the D1 shows a similar trend. To test the reliability of our inline HBT setup with different photon distributions, and exclude the presence of potential artifacts, we measured the autocorrelation of a single SPDC arm. **Figure 3d** shows $g_{ss}^{(2)}(\tau)$ of SPDC signal photons, measured using the two inline detectors without heralding. Although each SPDC arm individually exhibits thermal statistics, characterized by a peak at zero delay with $g_{ss}^{(2)}(0) = 2$, the measured data result from the convolution between the source correlation and the response function of the two detectors [44]. Given that the detectors' timing jitter is much larger than the coherence time of the signal photons, the convolution significantly broadens the peak, resulting in a flattened correlation curve. Overall, the results reported in Fig. 3d are consistent with this prediction and lack of any correlations indicated absence of spurious crosstalk between the two inline detectors. Finally, the conditional autocorrelation function $g_c^{(2)}(\tau)$ between two inline SNSPDs is presented in **Fig. 3e**. A clear antibunching dip is present at zero delay, with $g_c^{(2)}(0) = 0.019 \pm 0.001$, indicating high level of single-photon purity upon the heralding process. Again, the slight asymmetry of the curve is inherited from the cross-correlation function in Fig. 3c, which dominates the normalization factor of the conditional autocorrelation. These results are in good agreement with previous reports on SPDC-based sources [46, 47], attesting the reliability of our integrated architecture for characterizing quantum correlations.

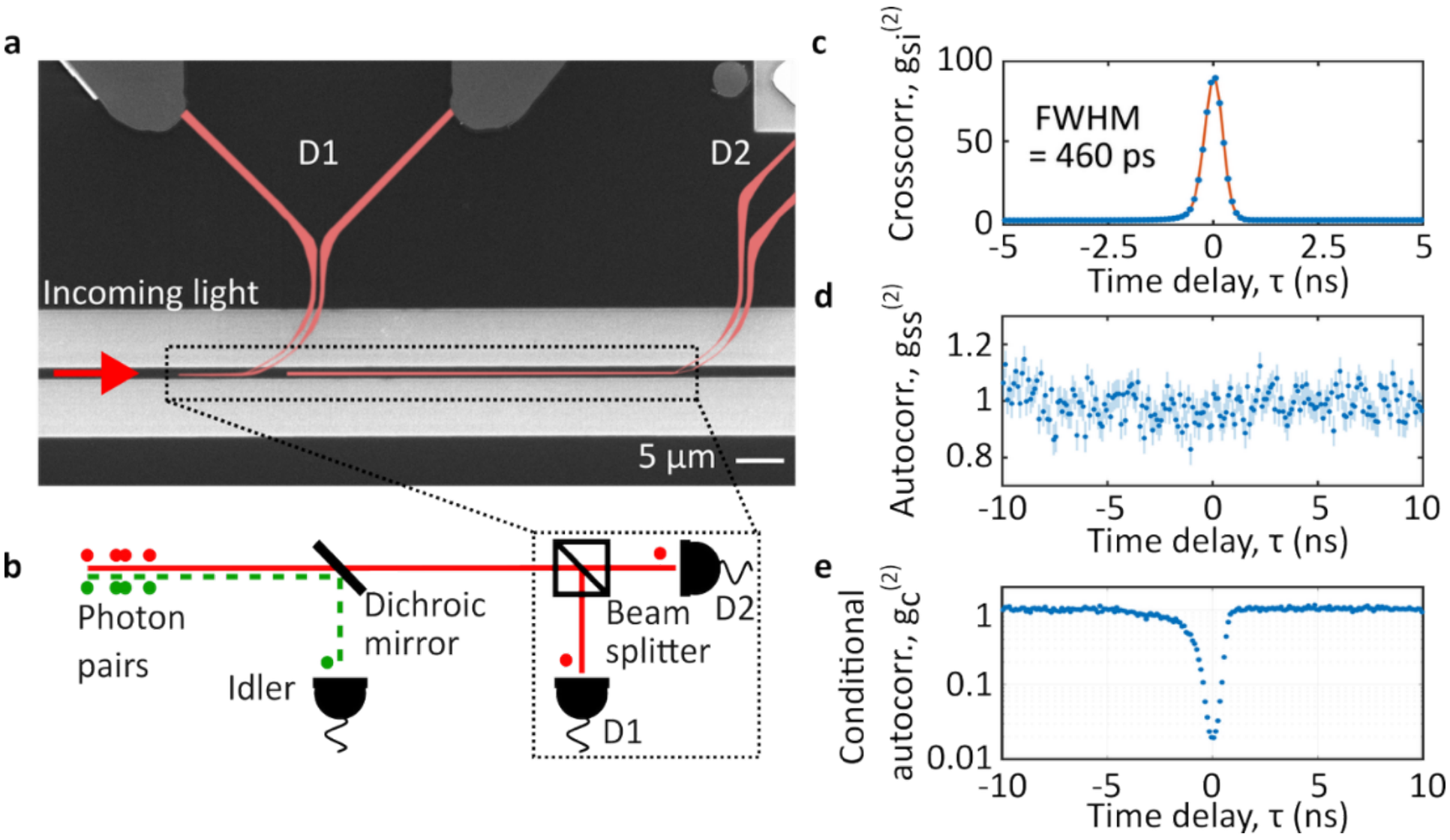

**Figure 3. Inline Hanbury Brown and Twiss setup for quantum correlation measurements.** (a) Microscope image of the inline HBT setup consisting of two SNSPDs (D1 and D2, false coloured) along a single waveguide. (b) Schematic of conditional autocorrelation setup for heralded single photons. Dashed boxes in (a) and (b) indicate the HBT setup. (c) Signal-idler cross correlation measurement (dots) and ex-Gaussian fitting (line). (d) Signal autocorrelation measurement without heralding. (e) Signal autocorrelation measurement conditional to heralding events.

As another application of inline quantum measurements, we implemented photon number resolution by collinear spatial multiplexing. PNR detectors are relevant for photonic quantum computing [48, 49], quantum communications [50], and quantum state preparation [51], where integrated approaches are being explored to improve scalability. Although SNSPDs exhibit some intrinsic photon-number discrimination capabilities [52], achieving practical PNR detectors with conventional SNSPD designs requires extremely low jitter and slow recovery dynamics [8], which can limit their applicability in high-speed circuits. Conversely, spatial multiplexing can relax these stringent temporal constraints, enabling efficient and scalable photon-number discrimination without compromising speed [53].

To experimentally validate the PNR capability of our inline detection scheme (**Fig. 4**), we used a pulsed femtosecond laser attenuated to the single-photon regime, with expected Poissonian statistics. We varied the average photon number per pulse ($\bar{n}$) in the range of 0.01 to 3 by adjusting the laser attenuation level. **Figure 4a** reports the count rate of each detector ($C_1$, $C_2$) together with the coincidence count rate ($C_{12}$) as a function of $\bar{n}$. The detectors exhibit a linear response up to $\bar{n} \sim 1$ photon per pulse (lines in Fig.4a). The sublinear response observed at higher $\bar{n}$ is attributed to a reduction in the effective detector efficiency arising from increased noise counts, e.g. those triggered by uncorrelated scattered photons (uncorrelated events increase with $\bar{n}$ while dark counts are negligible, see Supplementary Information, S6). This contribution overpowers the typical superlinear response induced by multiphoton contributions [54].

The overall probabilities of zero ($P_0$), one ($P_1$), or two ($P_2$ ) detection events per pulse (**Figs. 4b-d**) were extracted from the raw experimental data of Fig. 4a. These probabilities were compared with a theoretical model assuming Poissonian input statistics with average photon number $\bar{n}$. Under this assumption, the probability of no-click for each of the two nanowires is $P_0^{(1,2)} = e^{-\bar{n}\eta_{1,2}}$, where $\eta_{1,2}$ is the first/second nanowire detection efficiency. The overall probabilities $P_0$, $P_1$ and $P_2$ shown by the solid lines in Figs. 4b-d are straightforwardly derived from $P_0^{(1,2)}$ (see Methods). Experimental and theoretical data closely match each other, validating the accuracy of the inline architecture for resolving photon-number states.

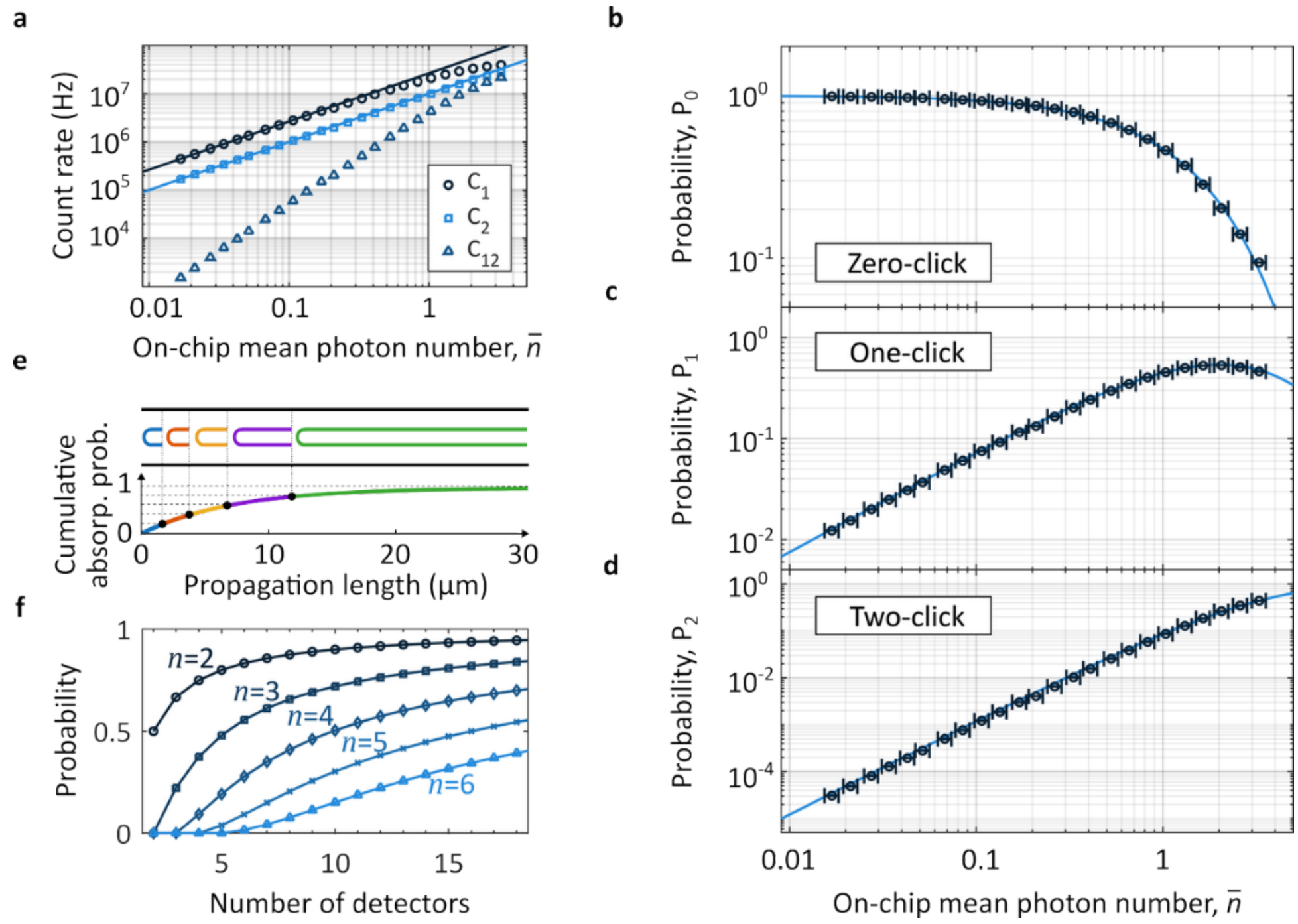

**Figure 4. Photon number resolution with inline detectors.** (a) Detectors count rate ($C_1$, $C_2$) and coincidences rate ($C_{12}$) at different on-chip mean photon numbers. The lines show the linear trend and are fitted on $C_1(C_2)$ at low mean photon number region. (b-d) Experimental data (dots) and fitting (lines) of zero-, one- and two-click probabilities as function of the mean photon number. Horizontal error-bars are based on the uncertainty estimated during the detection efficiency calibration (Supplementary Information, S4 and S5). Vertical error bars are within the marker size. (e) Schematic of five inline SNSPDs designed for equally distributed photon absorption. Graph shows cumulative probability of photon absorption across all the nanowires. (f) Fidelity of $n$-photon Fock state ($n$ = 2, 3, 4, 5 and 6) reconstruction as function of number of inline detectors.

Lastly, let us discuss the scalability of inline detection for larger photon-number states. By properly adjusting the length of multiple nanowires, the inline configuration can accommodate a larger number of detectors while maintaining balanced probability of photon detection. **Figure 4e** exemplifies this approach for a configuration with five nanowires, where the length of each nanowire is adjusted to absorb exactly 20% of the incoming light. **Figure 4f** represents the measurement fidelity (probability of correct reconstruction) of a $n$-photon Fock state as function of the number of inline nanowires [55]. As the number of nanowires must be large to accurately reconstruct high photon-number states, our inline approach represents a scalable approach to achieve high reconstruction fidelity within a compact footprint. Note that the ultimate scaling limit is set by the minimum absorption achievable in the first inline SNSPD. This is approximately 5.5% for our configuration, corresponding to a maximum of 18 nanowires, and could be further increased for example by nanowire composition, geometry, or coupling strength to the waveguide mode.

### 3. Discussion and conclusion

In this work, we introduced and formalized the concept of inline SNSPDs with distributed absorption, enabling quantum measurements in elementary photonic circuits with compact footprints. We developed an integration approach based on photonic bound states in the continuum to reduce scattering losses compared to conventional Si photonics waveguides, achieving competitive figures of merit for waveguide-integrated SNSPDs: on-chip detection efficiencies above 80%, timing recovery below 2 ns, and intrinsic jitter under 70 ps. Using this platform, we demonstrated two inline SNSPDs with balanced absorption for measuring photon correlations and statistics within a compact device footprint of 60 × 6 $\mu m^2$.

The compactness and scalability of the inline architecture are particularly evident in applications requiring a large number of detectors, such as photon-number resolution via spatial multiplexing. In conventional schemes, multiplexing is implemented with a mesh of beamsplitters, where detectors terminate the multiplexed waveguides [56]. This approach requires a total footprint of

$$A_{conv} \sim W_{BS} n \times (L_{BS} \log_2 n + L_{SNSPD}),$$

where $n$ is the number of detectors, $W_{BS}$ and $L_{BS}$ are the width and length of a single beamsplitter, and $L_{SNSPD}$ is the length of an integrated SNSPD. For example, with $n = 128$ detectors and conventional waveguide beamsplitters of dimensions $W_{BS} \times L_{BS} \approx 6 \times 70\ \mu m^2$ [57], the resulting PNR detector array occupies an area of $768 \times 510\ \mu m^2$. By contrast, in the inline implementation, $m$ detectors with equally distributed absorption terminate each waveguide, therefore resulting in a footprint of

$$A_{inline} \sim W_{BS} \frac{n}{m} \times \left( L_{BS} \log_2 \frac{n}{m} + m \langle L_{SNSPD} \rangle \right),$$

where $\langle L_{SNSPD} \rangle$ is an average length of inline SNSPDs. For $m = 18$ inline detectors per waveguide (see discussion above) and BIC beamsplitters with $W_{BS} \times L_{BS} \approx 18 \times 90\ \mu m^2$, the PNR detector footprint is reduced to $\approx 128 \times 309\ \mu m^2$ – a reduction of almost 10 times.

Compared to cascaded detector geometries that integrate SNSPDs with expanded, multi-mode waveguides [18], our inline architecture preserves single-mode operation. This aspect is particularly useful in quantum protocols that require weak or partial measurements along the light propagation path. For instance, inline measurement schemes are capable of conditional operations, such as multiphoton subtraction [58], without the need for beamsplitters. Similarly, multiple precisely placed and partially absorbing detectors can be employed to directly probe the quantum light states inside a circuit for direct monitoring of circuit operation [59].

Single-mode operation is also critical for measurement schemes relying upon light interference, including stationary-wave integrated Fourier-transform spectrometers (SWIFTS) [60] and for photon-number-resolving detectors [61]. In these devices, measurements are performed on standing waves formed by the interference of two counter-propagating traveling waves, and they require the integration of tens of equidistant detectors along a single waveguide. The small width (below 100 nm) of SNSPDs makes them ideally suited for probing standing-wave patterns, while the inline architecture offers a scalable, low-loss platform for implementing detector arrays with both high spatial precision and integration density.

Although this work has primarily focused on quantum measurements, the entire photonic circuit, including modulators and ring resonators, can be realized within the BIC platform [29]. In practice, however, the technologies for generating and manipulating quantum states of light are more mature and widely adopted in conventional platforms, such as silicon or silicon nitride waveguides. To fully exploit the advantages of inline measurements offered by the BIC platform, it will therefore be important to develop efficient integration strategies with these conventional platforms. Such integration can be achieved, for example, via transition tapers. As a concrete demonstration, in Section S7 of the Supplementary Information we show that a short (40 µm) linear taper enables coupling between a 500 nm ridge waveguide and a 1.55 µm BIC waveguide with sub-dB insertion losses.

In conclusion, we demonstrated platform of inline SNSPDs with distributed absorption as a compact and versatile route for implementing complex quantum measurements within a single optical waveguide, representing a promising pathway toward large-scale, multifunctional quantum photonic circuits.

## 4. Methods

### *4.1 Fabrication*

The integrated detectors were fabricated on a commercial SOI substrate (Wafer Pro, silicon thickness 220 nm, silicon dioxide thickness ~ 3000 nm). First, a 7 nm NbTiN film was deposited with magnetron co-sputtering with a process discussed in Ref. [35]. This film deposition was followed by three lithography steps to define alignment markers, nanowire and photonic circuit. Alignment markers and electrical contact pads were patterned with photolithography exposure of AZ5214E photoresist. A 4 nm layer of chromium and 96 nm layer of gold were deposited on the exposed chip via thermal evaporation (HV-experimentation system UNIVEX 250) and the excess of material was removed via lift-off in acetone. Secondly, the 60 nm wide nanowire was defined with a 30 kV electron-beam lithography (negative resist, AR-N 7520.073). The pattern was etched with a 40:1 mixture of $CF_4$ and oxygen by reactive ion etching (RIE; Oxford PlasmaLab 80). The addition of a small quantity of oxygen prevented micromasking and ensured full removal of the superconductor, which otherwise would increase the propagation losses. Finally, the photonic circuit was fabricated: a 200-nm-thick resist (CSAR 62, AR-P 6200.02) was spincoated on the chip and then exposed with 30 kV e-beam lithography. For experiments verifying the BIC condition and estimating propagation losses, devices were fabricated using only the final lithography step, omitting nanowire patterning.

*4.2 Photonic circuit characterization*

The BIC resonance and waveguide propagation losses were measured at room temperature using a custom-made photonic probe station with two bare fibres on a 6-axis stage. We used a continuous wave laser at 1550 nm (Santec, TSL-570) and a power meter (Thorlabs S154C) for these measurements. The fibres were aligned by maximizing the power transmitted through the circuit. A manual fiberised polarization controller was used to adjust the light polarization in the fibre. To characterize the nanowire absorption coefficient, we used the 4-port waveguide structure, shown in Fig. 3a. Light was delivered through an 8-deg fibre array (Meisu Optics) with 127 µm pitch, which matched the spacing between waveguide couplers on the chip.

*4.3 Cryogenic set-up*

Integrated detectors were characterized in a closed-cycle Grifford - McMahon (GM) cryostat with a base temperature of 3.5 K (DRY ICE 3 K, Ice Oxford). Optical input was provided through an 8-deg fibre array (Meisu Optics) mounted on a 4-axis stack of cryogenic nanopositioners (Attocube, ANP series). Fibre-to-chip alignment was ensured by maximizing the transmission between the two outer fibres and the reference loop 1-4 in Fig.3b. The 127 µm pitch between adjacent couplers matched the spacing of the fibre array, ensuring proper alignment of the central couplers. During SNSPD measurements, light in the reference loop was switched off to isolate the signal from the inline detectors. Flexible cryogenics cables (Delft Circuits Cri/oFlex) were used for electrical connections between the coldplate and room temperature. The detector was biased through a bias-tee (Minicircuits ZB85-12G-S +) with a low-noise current source (Yokogawa GS200) without additional filtering. The output RF signal was amplified by two room-temperature low-noise amplifiers (Minicircuits ZFL-1000LN +) and the pulses were recorded with a time tagger (ID Quantique, ID1000). The pulse shape was recorded with a 1 GHz Oscilloscope (LeCroy Wavesurfer 104MXs-B).

*4.4 On-chip detection efficiency characterization*

The on-chip detection efficiency was estimated as $\frac{LCR - DCR}{\bar{\eta}_c \cdot \Phi}$, where $LCR$ is the count rate under laser illumination, $DCR$ is the dark count rate, $\bar{\eta}_c$ accounts for coupler efficiency and waveguide propagation losses, and Φ is the photon flux delivered to the chip. The LCR was measured by sending CW laser light attenuated below single-photon level with fixed attenuators. The DCR was measured in the same alignment condition, but with the light off. An external calibrated power-meter, placed before the attenuators, was used to estimate the photon flux Φ that is reaching the chip during the LCR measurement. The coupling efficiency $\eta_c$ was estimated from the average transmission of four reference circuits equivalent to the one represented in Fig. 3a, but without the nanowire.

*4.5 Jitter measurement*

To measure the detector jitter (Fig. 3f), the device was irradiated with a femtosecond pulsed laser at 1550 nm (Calmar Laser, Mendocino, pulse width <500 fs, 50 MHz repetition rate). We acquired the interarrival time histogram between the laser trigger and the detector output signals. The measured system jitter, $\sigma_{\mathrm{FWHM}}$, was determined from the full-width half-maximum of the normalized histogram gaussian fitting.

This value contains the contribution of several sources: an electronic jitter ($\sigma_{electr}$) arising from the noise in the electronic readout and amplification system, a setup component ($\sigma_{setup}$) that includes contributions of the laser and photon counting module, and a component intrinsic to the detector and ($\sigma_{intr}$). Under the assumption that the contributions are independent, the system jitter can be written as

$$\sigma_{FWHM}^2 = \sigma_{electr}^2 + \sigma_{setup}^2 + \sigma_{intr}^2$$

The noise-induced jitter was estimated from the standard deviation of the electronic signal ($\sigma_{noise}$) and the slew-rate of the pulse rising edge (SR) as

$$\sigma_{electr} = 2 \cdot \sqrt{(2ln2)} \cdot \frac{\sigma_{noise}}{SR} = 29.5 \text{ ps}$$

The setup contribution was $\sigma_{setup} < 4.5$ ps, mainly due to the counter internal electronics and the laser trigger signal.

*4.6 Conditional autocorrelation measurement*

Photon pairs at 810 nm and 1550 nm were generated with collinear SPDC by pumping a PPKTP crystal with a CW 532 nm laser. Two dichroic mirrors were used to filter out the pump excess after the photon pairs generation, followed by a third dichroic to separate the signal from the idler. The idler photon (810 nm) was sent to a Si-SPAD (Excelitas SPCM-AQRH-14-FC, jitter > 350 ps) for the heralding process and the signal photon (1550 nm) was sent to the photonic chip with the inline HBT setup for the autocorrelation measurement. To evaluate the performance of the heralded single-photon source we performed a conditional correlation measurement, defined as

$$g_c^{(2)}(t_{s1}, t_{s2}|t_i) \equiv \frac{\langle \hat{E}_s^{\dagger}(t_{s1})\hat{E}_s^{\dagger}(t_{s2})\hat{E}_s(t_{s2})\hat{E}_s(t_{s1})\rangle_{pm}}{\langle \hat{E}_s^{\dagger}(t_{s1})\hat{E}_s(t_{s1})\rangle_{pm}\langle \hat{E}_s^{\dagger}(t_{s2})\hat{E}_s(t_{s2})\rangle_{pm}},$$

where the $\langle\cdot\rangle_{pm}$ indicates the average over the heralded events[20, 43].

Experimentally, the function was estimated as

$$g_c^{(2)}(\tau) = \frac{N_{is_1s_2}(0,\tau|0)N_i(0)}{N_{is_1}(0)N_{is_2}(\tau)}.$$

$N_{is_1s_2}(0,\tau|0)$ is the triple coincidence rate between the idler detection at $t_i = 0$, the signal 1 detection at $t_{s_1} = \pm\frac{w_{coinc}}{2}$ and the signal 2 detection at $t_{s_2} = \tau \pm \frac{w_{bin}}{2}$. The heralding coincidence window ($w_{coinc}$) was 1 ns, and the bin-width ($w_{bin}$) is 100 ps. $N_i(0)$ is the average rate of the idler signal and $N_{is_1}(0)$ ($N_{is_2}(\tau)$) is the double coincidence rate between idler and signal 1 (signal 2). Data were acquired for 12 h. The relative uncertainty for each measured rate was estimated as $u_{r(N)} = \frac{\sqrt{N}}{N} = \frac{1}{\sqrt{N}}$ due to the light shot noise.

*4.7 Photon number resolution measurement*

Photon-number statistics were measured using the same experimental setup employed for jitter characterization. Both detectors were biased at 85% of their respective critical currents, showing a dark counts rate ~ 3 kHz, across the two detectors. To minimize contributions from dark counts, only detection events occurring within a ±1 ns timing window around the laser trigger pulse were counted. Within this time window, the probability of dark count event was negligible, $\sim 6 \cdot 10^{-6}$. Each data point was acquired for 10 s, corresponding to more than $5 \cdot 10^{8}$ laser triggering events. The laser intensity was adjusted with a variable attenuator and the on-chip mean photon number per pulse was varied between 0.01 and 3. A power meter was used to monitor the laser intensity during the measurement and the average photon number per pulse was estimated as $\bar{n} = \frac{P_{PM} \cdot A_{losses}}{C_{Tr} \cdot E_{ph}}$, where $P_{PM}$ is the power measured at the powermeter, $A_{losses}$ includes channel losses, attenuation and coupling efficiency, $C_{Tr}$ is the laser repetition rate and $E_{ph}$ is the energy of a photon. The relative uncertainty of $\bar{n}$ is the same as the photon flux used in the on-chip detection efficiency estimation (Supplementary Information, S4 and S5). From the recorded data, experimental two-, one- and zero-click probabilities were respectively calculated as

$$P_2 = C_{12}/C_{Tr},\ P_1 = (C_1 + C_2 - 2C_{12})/C_{Tr} \text{ and } P_0 = 1 - (P_1 + P_2).$$

Experimental data were fitted with a theoretical model, assuming input Poissonian statistics and constant detector efficiency. Under these assumptions, the probability of no-click for each detector is $P_0^{(1)} = e^{-\bar{n}\eta_1}$ and $P_0^{(2)} = e^{-\bar{n}\eta_2}$, with detector efficiencies $\eta_1$ and $\eta_2$. Theoretical total zero-, one- and two-click probabilities were then calculated as

$$P_0 = P_0^{(1)} \cdot P_0^{(2)},\ P_1 = P_0^{(1)} \cdot \left(1 - P_0^{(2)}\right) + \left(1 - P_0^{(1)}\right) \cdot P_0^{(2)}, \text{ and } P_2 = \left(1 - P_0^{(2)}\right) \cdot \left(1 - P_0^{(1)}\right).$$

**Acknowledgments**

We are grateful to Harish N. S. Krishnamoorthy for the valuable discussions at the inception of this work and Ruixiang Guo for the help in setting up and characterizing the SPDC source.

**Funding**

Research was supported by the Singapore National Research Foundation through the Quantum Engineering Programme (QEP-P1 and NRF2021-QEP2-01-P01) and the National Centre for Advanced Integrated Photonics (NRF-MSG-NCAIP), and by the Ministry of Education through the NTUitive GAP Fund (NGF-2024-16-015).

**Contribution Statement**

F.M. performed numerical simulations, device nanofabrication and testing. S.D. optimized superconducting film deposition. D.M.Z.K., under supervision of C.K., developed the SPDC source and assisted with correlation measurements. M.S. contributed to the establishment of detector characterization methodologies. F.M., A.N.V. and C.S. analysed the data and drafted the manuscript. C.S. supervised the work. All authors have accepted responsibility for the entire content of this manuscript and approved its submission.

**Disclosures**

The authors report no conflict of interests.

**Supplementary document**

See Supplement 1 for supporting content.